\documentclass[doublecol]{epl2} 

\usepackage{amsmath}
\usepackage{amsfonts}
\usepackage{amssymb}
\usepackage{graphicx}
\usepackage{dcolumn}
\usepackage{bm}
\usepackage{braket}
\usepackage{amsthm}
\usepackage{xr-hyper}
\usepackage{hyperref}
\usepackage{bookmark} 
\usepackage{caption}
\usepackage[pagewise, switch, displaymath, mathlines, columnwise]{lineno}
\usepackage{cuted} 

\title{Semiclassical kinetic theory for systems with non-trivial quantum geometry
and the expectation value of physical quantities}

\shorttitle{Semiclassical kinetic theory for systems with non-trivial quantum geometry} 

\author{Thierry Valet \and Roberto Raimondi}

\institute{                    
  Dipartimento di Matematica e Fisica,  Universit\`a  Roma Tre, Via della Vasca Navale 84, 00146 Roma, Italy
}

\abstract{
Starting from the Keldysh theory, for a general low energy $N$-band Hamiltonian in the clean limit, we perform a manifestly $\smash{U(1) \times SU(N)}$ gauge invariant semiclassical expansion. A generalized Berry curvature tensor is shown to control a redistribution of spectral weights. New expressions for certain physical quantities ensue, establishing the limits of a previously proposed correction to the density of states. In the two-band case, we derive a completely general semiclassical kinetic theory including all $O(\hbar)$ quantum corrections. As an application, we show how one can recover, out of a single simple calculation, the chiral anomaly, intrinsic anomalous Hall conductivity and chiral magnetic effect, in all generality.  The demonstrated flexibility and efficiency of our formalism derives from the insulation it provides from the underlying complexity of the quantum kinetics, notwithstanding its rigorous connection to this deeper level.}

\begin{document}

\maketitle

{\em Introduction--} The wave packet approach to derive semiclassical equations of motion \cite{sundaram:1999,xiao:2010} has been very successful modeling leading order quantum geometric corrections to transport phenomena \cite{jungwirth:2002,matsumoto:2011,cheng:2016,zhang:2019,son:2013,yang:2015,johansson:2019,wang:2021}. However, and despite previous attempts \cite{shindou:2008,wong:2011,wickles:2013}, a rigorous justification of this approach in multiband situations, as a controlled approximation of a deeper level of quantum theory, is still lacking in our opinion. For instance, as shown in the pioneering work by Wong and Tserkovniak \cite{wong:2011} for a general two-band system, a semiclassical expansion of the density matrix evolution equation uncovers interband couplings not found in the wave packet approach. Combining techniques borrowed from high energy physics \cite{vasak:1986,huang:2018}, with a manifestly $SU(N)$ gauge invariant formulation of the quantum geometry \cite{graf:2021}, we report here about a controlled $\hbar$ expansion of the gauge invariant quantum kinetic theory for a general low energy $N$ band Hamiltonian in the clean limit. A generalization of the Berry curvature \cite{berry:1989} tensor is found to control newly uncovered spectral weight transfers, which resolve the  apparent inconsistencies between the wave packet and density matrix approaches. However, a modification to the density of states, previously introduced in the wave packet approach \cite{xiao:2005}, is demonstrated inadequate and new expressions for the expectation value of general observables are derived. In the two-band case, the power of our gauge invariant semiclassical kinetic theory is illustrated with a rigorous and general, yet strikingly intuitive, derivation of the chiral anomaly, intrinsic anomalous Hall effect and chiral magnetic effect.   

{\em Gauge invariant quantum kinetic theory--} Even though we develop a non-relativistic kinetic theory, we choose to formulate it initially using Lorentz covariant notations, which greatly facilitates the derivation. Hence, the phase space variables are defined as the spacetime position, $\smash{{x \equiv x^{\mu} \equiv (c t, x^ i) \equiv (c t, {\bf x})}}$, and the kinetic energy-momentum, $\smash{{p \equiv p^{\mu} \equiv (\varepsilon / c, p^i) \equiv (\varepsilon / c, {\bf p})}}$, with $c$ being the velocity of light. The lowering or raising of spacetime indices results, in a standard way, from the action of the Minkowski metric tensor, whose signature is chosen to be $(+, -, -, -)$, and Einstein summation convention is used throughout. We introduce the band structure at a semi-microscopic level, in the form of a general multi-band low energy Hamiltonian \cite{luttinger:1951, luttinger:1955, kohn:1959} {\em i.e.}, an energy matrix  in phase space, $\smash{{\hat h}(x,{\bf p})}$, acting on the Hilbert space of dimension $N$ of the internal degrees of freedom (spin, orbital). Some possible smooth spacetime perturbations of the system, to the exclusion of the electromagnetic field, are encoded in the explicit $x$ dependence. Such model Hamiltonian may results from a $k.p$ expansion, or from the continuum limit of a tight-binding model, possibly fitted to ab-initio band structure calculations. This illustrates the flexible nature of our approach, in its ability to equally handle archetypal ``toy'' Hamiltonian or more realistic material models. We also define $\smash{{\hat \lambda}(x,p) = \varepsilon {\hat \gamma}_{0} - {\hat h}(x,{\bf p})}$, with ${\hat \gamma}_0$ the identity matrix of dimension $N$. This central quantity can be recognized as the Wigner transform \cite{weyl:1927,wigner:1932} of the operator whose expectation value is a suitable Lagrangian density for the Schr\"odinger field. Hence, we will refer to it as the Lagrangian function and we will see that it is the actual quantity controlling the non-interacting out-of-equilibrium quantum statistical evolution. This evolution, for a many electron system, is most efficiently formulated in terms of real-time Green's functions \cite{schwinger:1961, kadanoff:1962, keldysh:1965}, a theoretical framework usually referred to as the Keldysh theory. As our goal is to derive a semiclassical transport theory, we naturally focus on a formulation of Keldysh theory in phase space \cite{mahan:1990, rammer:1986}. Furthermore, in order to enforce from the outset the manifest electromagnetic gauge invariance of the theory, one shall rely on a $U(1)$ gauge invariant generalization of the Wigner transformation \cite{stratonovich:1956,bialynicki:1977,serima:1986,vasak:1986,levanda:1994,muller:1999,levanda:2001}. In such context, the central quantity is the lesser Green's function in phase space {\em i.e.}, $\hat{g}_{<}(x,p)$, which encodes the out-of-equilibrium state of the system. This quantity is related to the equal time kinetic level {\em i.e.}, the density matrix level, through the gauge invariant Wigner function \cite{serima:1986}, $\smash{{\hat \rho}(x, {\bf p})}$, which is defined as the Wigner transform of the density matrix. It can be obtained from $\smash{{\hat g}_{<}}$ by energy integration \cite{levanda:1994} {\em i.e.}, $\smash{{\hat \rho}(x, {\bf p}) = \frac{1}{2 {\bf i} \pi} \int d \varepsilon \ {\hat g}_{<}(x, p)}$. The key property of the gauge invariant Wigner function \cite{hillery:1984,serima:1986} is that it yields the expectation value of any single particle observable, $\smash{{\hat O}}$, in the form of its local density, $\smash{\langle \rho_{o} \rangle(x)}$, as a weighted integral over momentum of its Weyl symbol $\smash{{\hat o}({\bf x}, {\bf p})}$:
\begin{equation} \label{eq:lett_wigner_avg}
\langle \rho_{o} \rangle(x) = \int \frac{dp}{(2 \pi \hbar)^d} \ \mathrm{Tr} \left[ {\hat \rho}(x, {\bf p}) \ {\hat o}({\bf x}, {\bf p}) \right] ,
\end{equation}
making finally contact with measurable physical quantities. Now that we have outlined the theoretical underpinning of our approach, we can state our concrete starting points, in the form of the governing differential equations in phase space for the Green's functions. For the lesser Green's function, and in respective order, we have the quantum kinetic equation (QKE), obtained from the difference between the left and right Dyson's equations, and the constraint equation (CE) obtained from the half sum:
\begin{subequations} \label{eq:lett_green_clean}
\begin{eqnarray} 
\Big[ \hat{\lambda} \stackrel{\star}{,} \hat{g}_{<} \Big]_{(-)}  & = & 0 , \label{eq:lett_qke_clean} \\
\frac{1}{2} \Big[ \hat{\lambda} \stackrel{\star}{,} \hat{g}_{<} \Big]_{(+)} & = & 0 , \label{eq:lett_ce_clean}
\end{eqnarray}
\end{subequations}
with the $(-)$ and $(+)$ subscripts indicating respectively a commutator and an anti-commutator. The QKE, if often stated for a single band and in a gauge dependent way \cite{rammer:1986,mahan:1990}, is well understood as a quantum precursor to the Boltzmann equation and quasiclassical transport equations. However, we want to emphasize here the equal importance of the CE, which has been so far almost universally overlooked in non-relativistic quantum kinetic theory applied to condensed matter systems. When performing a systematic order by order expansion in terms of $\hbar$, the CE provides indeed constraints on the structure of the lesser Green's function. This is systematically used in relativistic quantum kinetic theory \cite{vasak:1986, huang:2018}, and will be used extensively in this work.  We also need to consider the retarded (advanced) Green's function, which determines the quasiparticle  spectrum, and which obeys:
\begin{subequations}
\label{eq:lett_invariant_RA_clean} 
\begin{eqnarray}
\label{eq:lett_invariant_RA_qke_clean}
\Big[ \hat{\lambda} \stackrel{\star}{,} \hat{g}_{R(A)} \Big]_{(-)} & = & 0, \\ 
\label{eq:lett_invariant_RA_ce_clean}
\frac{1}{2} \Big[ \hat{\lambda} \stackrel{\scriptscriptstyle +}{\scriptscriptstyle (-)} {\bf i} 0^{+} {\hat \gamma}_0  \stackrel{\star}{,} \hat{g}_{R(A)} \Big]_{(+)} & = & {\hat \gamma}_0 ,  
\end{eqnarray}
\end{subequations}
with $\smash{{\bf i}^2 = -1}$. The star product, which is the proper gauge invariant generalization \cite{muller:1999,levanda:2001,onoda:2006,gorini:2010,shitade:2017}, in presence of an electromagnetic field, of the better known Moyal's product \cite{groenewold:1946, moyal:1949}, replaces the ordinary matrix product in Eqs.({\ref{eq:lett_green_clean}}-\ref{eq:lett_invariant_RA_clean}). While we have written here these equations in their non-interacting clean limit, the inclusion of disorder and/or interactions is possible by addition of suitable self-energies terms \cite{rammer:1986,rammer:2007}. We intend to explore the resulting interplay with quantum geometry in future works as it has been done within the density matrix context\cite{culcer:2017,sekine:2017,atencia:2022}. 
For further details on the formalism, we specifically refer the reader to \cite{rammer:2007} regarding the Keldysh theory and its semiclassical limit, and to \cite{onoda:2006, shitade:2017} and references herein, for the delicate questions raised by gauge invariance and the star product.

{\em Quantum geometry--} We adopt a manifestly $SU(N)$ gauge invariant formulation of quantum geometry \cite{graf:2021}. We introduce the eigenvalues $\smash{\epsilon^{n}(x, {\bf p})}$, assumed to be nondegenerate almost everywhere, and the associated eigenprojectors:
\begin{equation}
	{\hat P}_{n}(x, {\bf p})  = \frac{1}{N} {\hat \gamma}_0 + \frac{1}{2} {\bf b}_{n}(x, {\bf p}) \cdot \hat{\bm{\gamma}} ,
\end{equation}
of $\smash{{\hat h}}$, in which ${\hat{\bm{\gamma}} = (\hat{\gamma}_1, \ldots, \hat{\gamma}_{N^2 - 1})}$ is the vector of the traceless Hermitian generators of the $\mathfrak{su}(N)$ Lie algebra and with $\smash{n \in \lbrace 1, \ldots N \rbrace}$ the band index. The generators can be conveniently chosen to be the generalized Gell-Mann matrices \cite{bertlmann:2008}, which reduce into the Pauli matrices in the two-band case. The real vectors ${\bf b}_{n}$ are the generalized Bloch vectors, which can be computed from the sole knowledge of the energy matrix and its eigenvalues, and entirely determine the quantum geometric tensor \cite{graf:2021,hioe:1981}. By definition we have $\smash{{\hat h} = \sum_{n} \varepsilon^{n} {\hat P}_{n}}$, and in all generality:
\begin{equation}
{\hat \lambda} = \sum_{n} \lambda^{n} {\hat P}_{n} = \lambda^0 {\hat \gamma}_0 - \sum_{n} \frac{\Delta^{n}}{2} {\bf b}_{n} \cdot \hat{\bm{\gamma}} ,
\end{equation}
with $\smash{\lambda^{n} = \varepsilon - \varepsilon^{n}}$, $\smash{\lambda^0 = \varepsilon - \varepsilon^0}$, $\smash{\varepsilon^0 = \frac{1}{N} \sum_{n} \epsilon^{n}}$, and $\smash{\Delta^{n} = \varepsilon^{n} - \varepsilon^0}$. Similarly, we have a unique geometric decomposition of the Green's functions given by:
\begin{subequations}
\begin{eqnarray}
	& &{\hat g} = \sum_{n} g^{n} {\hat P}_{n} + \frac{1}{2} {\bm g}^{\bot} \cdot \hat{\bm{\gamma}} , \\
	& &\forall n, {\bf b}_{n} \cdot {\bm g}^{\bot} = 0 .
\end{eqnarray}
\end{subequations} 
Here ${\hat g}$ is standing for either ${\hat g}_{<}$ or ${\hat g}_{R(A)}$. The band projected part of ${\hat g}$ is then obtained as $\smash{g^{n} = \mathrm{Tr} ( {\hat g} {\hat P}_{n} )}$, while $\smash{{\bm g}^{\bot}}$ is the transverse (off-diagonal) part which relates to interband coherences. 

{\em Semiclassical expansion--} To constrain our theory at the semiclassical level, we introduce a formal expansion of the Green's functions in power of $\hbar$:
\begin{equation}
	{\hat g} =  {\hat g}^{(0)} +  {\hat g}^{(1)} + O(\hbar^2) = {\hat {\bar g}} + O(\hbar^2),
\end{equation}
with $\smash{{\hat g}^{(k)} \sim O(\hbar^k)}$ and \smash{${\hat {\bar g}}$} capturing all leading order quantum corrections. We simultaneously introduce the gradient expansion of the star product \cite{onoda:2006, shitade:2017}, which results from the expansion of the Abelian gauge links (Wilson lines) central to the definition of a U(1) gauge invariant Wigner transform. It can be expressed as:
\begin{equation}
\hat{\lambda} \star {\hat g} \! = \hat{\lambda} {\hat g} + {\bf i} \ \frac{\hbar}{2} \lbrace \hat{\lambda} , {\hat g} \rbrace - \frac{1}{2} \left( \frac{\hbar}{2} \right)^2 \lbrace \lbrace \hat{\lambda}, {\hat g} \rbrace \rbrace +  O(\hbar^3) ,
\end{equation}
with:
\begin{equation}
\lbrace \hat{\lambda} , {\hat g} \rbrace  =  \partial_p^{\mu} \hat{\lambda} \ D_{\mu} {\hat g} - D_{\mu} \hat{\lambda} \ \partial_p^{\mu} {\hat g} ,
\end{equation}
which can be interpreted as a $U(1)$ gauge invariant generalization of the Poisson bracket. This gauge invariant Poisson bracket is defined in terms of the gradient in kinetic energy-momentum $\smash{\partial^\mu_p \equiv (c \partial_\varepsilon, \partial^i_p) \equiv (c \partial_\varepsilon, - \partial_{\bf p})}$, and of a gauge invariant derivative:
\begin{equation} \label{eq:inv_D}
	D_{\mu} \equiv \partial_{\mu} - \frac{e}{2} F_{\mu \nu} \partial_{p}^{\nu} ,
\end{equation}
with \mbox{$e (< 0)$} the electron charge, $F_{\mu \nu}$ the electromagnetic field tensor, and $\smash{\partial_\mu \equiv (\frac{1}{c}\partial_t, \partial_i ) \equiv (\frac{1}{c}\partial_t, \partial_{\bf x})}$ the spacetime gradient. As this operator is certainly new to the reader, we want to stress that it shall not be confused with the more commonly defined gauge covariant derivative, involving the 4-vector potential. We shall also outline that this gauge invariant differential operator, as can be clearly seen in Eq.(\ref{eq:inv_D}), mixes spacetime and momentum derivatives in presence of a constant and uniform field tensor. This is a direct manifestation of the electromagnetic field as a deformation parameter of the symplectic geometry of the phase space associated with a charge particle \cite{karasev:2004}. As for the $O(\hbar^2)$ term, it can be found for instance in \cite{shitade:2017}, and exhausts all nonlinear quantum corrections up to quadratic order in a uniform and constant field strength. We are now in a position to perform an order by order expansion \cite{vasak:1986,huang:2018} and band projection of \mbox{Eqs.({\ref{eq:lett_green_clean}}-\ref{eq:lett_invariant_RA_clean}).} This is done by substituting into these equations the $\hbar$ expansions for the Green's functions and star product, and solving at each order for the transverse part as a function of the longitudinal components \cite{wong:2011}, allowing for the elimination of the former. A non-relativistic semiclassical kinetic theory is consistently obtained by truncating this procedure at the $O(\hbar^2)$ level.

{\em Berry curvature modifications to the electronic spectrum--} We first focus on the spectral weight function, which encodes the single particle excitations \cite{rammer:2007}. At the semiclassical level, it is given by $\smash{{\hat {\bar a}} = {\bf i}({\hat{\bar g}}_{R} - {\hat{\bar g}}_{A})}$. From Eqs.({\ref{eq:lett_invariant_RA_qke_clean}-{\ref{eq:lett_invariant_RA_ce_clean}) we obtain:
\begin{equation} \label{eq:lett_invariant_RA_ce_clean_bar}
\!\! {\bar \lambda}^{n}_{\stackrel{\scriptscriptstyle +}{\scriptscriptstyle (-)}} {\bar g}_{R(A)}^{n} + \frac{\hbar}{2} \sum_{m \neq n} \left( \Delta^{n} \! - \! \Delta^{m} \right) \Omega^{n}_{m}{} \ {\bar g}_{R(A)}^{m}  = 1 + O(\hbar^2),
\end{equation}
with $\smash{{\bar \lambda}^{n}_{\stackrel{\scriptscriptstyle +}{\scriptscriptstyle (-)}} = \left( \varepsilon \stackrel{\scriptscriptstyle +}{\scriptscriptstyle (-)} {\bf i} 0^{+} \right) - {\bar \varepsilon}^{n}}$, and:
\begin{equation} 
{\bar \varepsilon}^{n} = \varepsilon^{n} + \frac{\hbar}{2} \sum_{m \neq n} \left( \Delta^{n} \! - \! \Delta^{m} \right) \Omega^{n}_{m}{} ;\label{eq:lett_band_Nfirst}
\end{equation}
in which we have introduced $\smash{\Omega^{n}_{m}{} = \Omega^{n}_{m}{}^i{}_i{}}$, the contraction of the mixed sector in phase space of the tensor $\smash{\underline{\underline{\Omega}}^{n}_{m}{}}$, with:
\begin{eqnarray} 
\!\!\!\!\! \Omega^{n}_{m}{}^{\mu}{}_{\nu}{} = - \frac{1}{2} {\bf b}_{n} \cdot \left( \partial_p^\mu  {\bf b}_{m} \times D_{\nu} {\bf b}_{m} \right) . \label{eq:lett_mixed_gauge_berry}
\end{eqnarray}
The components of $\smash{\underline{\underline{\Omega}}^{n}_{m}{}}$ in the spacetime sector {\em i.e.}, $\smash{\Omega^{n}_{m}{}_{\mu}{}_{\nu}{}}$, and in the momentum sector {\em i.e.}, $\smash{\Omega^{n}_{m}{}^{\mu}{}^{\nu}{}}$, can be immediately inferred from Eq.(\ref{eq:lett_mixed_gauge_berry}) by obvious differential operator substitutions as dictated by the chosen notation. {\em We shall outline, and have to acknowledge, that we are here committing an abuse of the standard relativistically covariant notations used thorough this article.  In the case of the $\smash{\underline{\underline{\Omega}}^{n}_{m}{}}$ tensor, upper and lower spacetime indices are used as a compact notation to refer to all four phase space sectors of this tensor, in a unified and compact way.} It shall also be kept in mind, in relation with Eq.(\ref{eq:inv_D}) and the adjacent discussion, that in presence of an electromagnetic field the different phase space sectors of $\smash{\underline{\underline{\Omega}}^{n}_{m}{}}$ all acquires a mixed character, to the exception of the momentum sector. The cross product appearing in Eq.(\ref{eq:lett_mixed_gauge_berry}) is the vector product over $\smash{\mathbb{R}^{N^2-1}}$ induced by the anti-symmetric structure constants of the $\mathfrak{su}(N)$ Lie algebra \cite{mallesh:1997, graf:2021}.  It is immediate to recognize $\smash{\underline{\underline{\Omega}}^{n}_{m}{}}$ as a generalized Berry curvature tensor in phase space, in presence of an electromagnetic field. Our Eq.(\ref{eq:lett_mixed_gauge_berry}) is providing a manifestly $\smash{U(1) \times SU(N)}$ gauge invariant definition for this tensor. When $n = m$, $\smash{\underline{\underline{\Omega}}^{n}_{n}}$ is related to the Abelian Berry curvature tensor for the band $n$, see Eq.(35) in Ref.\cite{graf:2021}. When \mbox{$n \neq m$}, $\smash{\underline{\underline{\Omega}}^{n}_{m}{}}$ relates to the interband Berry curvature previously introduced in the theories of the circular photogalvanic effect \cite{xu:2018,kim:2019} and nonlinear Hall effect \cite{okyay:2021}. One can demonstrate that $\smash{\underline{\underline{\Omega}}^{n}_{m}{}}$ verifies the sum rule $\smash{\sum_{m \neq n} \underline{\underline{\Omega}}^{m}_{n}{} = - \underline{\underline{\Omega}}^{n}_{n}{}}$, as well as $\smash{\sum_{n} \underline{\underline{\Omega}}^{n}_{n}{} = 0}$ \cite{xiao:2010}. After solving Eq.(\ref{eq:lett_invariant_RA_ce_clean_bar}) for ${\bar g}_{R(A)}$, we obtain:
\begin{eqnarray}
{\hat {\bar a}} = & 2 \pi & \sum_{n} \Bigg[\left( 1 - \frac{\hbar}{2} \Omega^{n}_{n} \right) {\hat P}_{n} + \frac{\hbar}{2} \sum_{m \neq n} \Omega^{m}_{n} {\hat P}_{m} \Bigg] \delta(\varepsilon - {\bar \varepsilon}^{n})  \nonumber \\
& + & \frac{1}{2} \ {\bar {\bm a}}^{\bot} \cdot {\hat {\bm \gamma}} , \label{eq:lett_Nmatrix_spectrum}
\end{eqnarray}
and taking the trace:
\begin{equation} \label{eq:lett_Nmatrix_diag_spectrum}
\mathrm{Tr} \left( {\hat {\bar a}} \right) = 2 \pi \sum_{n} \left( 1 - \hbar \Omega^{n}_{n} \right)  \delta(\varepsilon - {\bar \varepsilon}^{n}).
\end{equation}
{\em Equations (\ref{eq:lett_band_Nfirst}-\ref{eq:lett_Nmatrix_diag_spectrum}) constitute our first significant result}. We see that the elementary excitations of the system remain, as expected, perfectly defined quasiparticles in $N$ separate energy bands. We also observe a change in the quasiparticle energies, given by Eq.(\ref{eq:lett_band_Nfirst}). It is easy to verify that it is identical to the energy correction first obtained by Sundaram and Niu with the wave packet approach \cite{sundaram:1999,xiao:2010}. We recall that the (in general affine) dependence of this energy correction with an external magnetic field allows to define the orbital magnetic moment carried by the quasiparticles \cite{xiao:2010}. We also find that the quasiparticles acquire Berry curvature corrected spectral weights, as revealed by Eq.(\ref{eq:lett_Nmatrix_spectrum}). We see that a quasiparticle belonging to the energy band $n$ carries a spectral weight changed (from unity) to $\smash{Z^{n}_{n} = ( 1 - \frac{\hbar}{2} \Omega^{n}_{n})}$, respectively to the local eigenvector $| n(x,p) \rangle$ of $\hat{h}(x,p)$. Simultaneously it carries nonzero spectral weights, $\smash{Z^{m}_{n} =  \frac{\hbar}{2} \Omega^{m}_{n}{}}$, respectively to the other eigenvectors $| m(x,p) \rangle$. This ultimately results in quasiparticles carrying a total spectral weight which becomes band dependent {\em i.e.}, $\smash{Z^{n}_{(tot)} = Z^{n}_{n}{} + \sum_{m \neq n} Z^{m}_{n}{} = ( 1 - \hbar \Omega^{n}_{n} )}$, and we have $\smash{\sum_{n} Z^{n}_{(tot)} = N}$ as expected. This spectral weight redistribution is a significant new finding of this work. While a change to the total spectral weight $Z^{n}_{(tot)}$ was previously found at the Keldysh theory level \cite{shindou:2008} and the density matrix theory level \cite{sekine:2017}, the existence of Berry curvature induced interband spectral weight transfers is to the best of our knowledge uncovered here for the first time. This has immediate consequences already in equilibrium, which may exist under a constant and uniform magnetic field, ${\bf B}$. From the definitions of Eq.(\ref{eq:lett_mixed_gauge_berry}) and Eq.(\ref{eq:inv_D}), we obtain \mbox{$\smash{\Omega^{(eq)}{}^{n}_{m}{}\left( {\bf B} \right) = e {\bf B} \cdot {\bm \Omega}^{n}_{m}{}}$,} in which we have introduced the generalized (axial) Berry curvature vector in momentum space defined by $\smash{({\bm \Omega}^{n}_{m}{})_k = \frac{1}{2} \ \epsilon_{i j k} \Omega^{n}_{m}{}^{i j}{}}$, with $\smash{\epsilon_{i j k}}$ the Levi-Civita symbol. We also recall the fluctuation-dissipation relation for ${\hat g}_{<}$ \cite{rammer:1986} {\em i.e.}, $\smash{{\hat g}_{<}^{(eq)} = {\bf i} \ {\hat a}^{(eq)} n_F(\varepsilon)}$, with $\smash{n_F}$ the Fermi-Dirac distribution. Then, from Eq.(\ref{eq:lett_Nmatrix_spectrum}), we immediately obtain:
\begin{eqnarray} \label{eq:lett_less_eq}
\!\!\!\!\!\! {\bar g}^{(eq) n}_{<} = 2 {\bf i} \pi \bigg[ & & \!\!\!\!\!\!\!\!\! \left( 1 - \frac{e \hbar}{2} {\bf B} \cdot {\bm \Omega^{n}_{n}}\right) \delta(\varepsilon - {\bar \varepsilon}^{n})  \nonumber \\ 
 & + &  \frac{e \hbar}{2} \sum_{m \neq n} {\bf B} \cdot {\bm \Omega}^{n}_{m}{} \ \delta(\varepsilon - {\bar \varepsilon}^{m}) \bigg] n_F(\varepsilon) ,
\end{eqnarray}
which, after energy integration and applying Eq.(\ref{eq:lett_wigner_avg}), gives us the equilibrium electron density as:
\begin{equation}
n_{e}^{(eq)} = \sum_{n} \int \frac{dp}{(2 \pi \hbar)^d} ( 1 - e \hbar {\bf B} \cdot {\bm \Omega^{n}_{n}} ) n_F({\bar \varepsilon}^{n}) .
\end{equation}
This expression for the electron density is identical to the prediction of the wave packet approach {\em i.e.}, Eq.(7) of Ref.\cite{xiao:2005}, leading for instance to the same expression for the intrinsic Hall conductivity as derived from the Streda formula \cite{streda:1982}. However, in our theory, the prefactor $\smash{Z^{n}_{c}=( 1 - e \hbar {\bf B} \cdot {\bm \Omega^{n}_{n}})}$ clearly originates from the Berry curvature corrected total spectral weight, $Z^{n}_{(tot)}$. Conversely, within the wave packet approach, this factor is introduced as a Berry curvature correction to the density of states in phase space \cite{xiao:2005} in order to enforce the assumed validity of the Liouville theorem in each band considered separately. We are not the first to point at this distinction \cite{sekine:2017}. However, as we will see in the following, the newly uncovered interband spectral weight redistribution implies more than just an interpretative change. This has in fact physical consequences when it comes to computing the expectation value of general observables.

{\em Gauge invariant semiclassical kinetic theory--} We consider now a general out-of-equilibrium situation. The band projection of the CE gives us:
\begin{equation} \label{eq:lett_invariant_lesser_ce_clean_bar}
{\bar \lambda}^{n} {\bar g}_{<}^{n} + \frac{\hbar}{2} \sum_{m \neq n} \left( \Delta^{n} \! - \! \Delta^{m} \right) \Omega^{n}_{m}{} \ {\bar g}_{<}^{m} = O(\hbar^2) ,
\end{equation}
for which a general solution in the sense of distribution, and with $O(\hbar)$ accuracy, is \mbox{$\smash{{\bar g}^{n}_{<} = \delta({\bar \lambda}^{n}) \mathcal{F}^{n} + \frac{\hbar}{2} \sum_{m \neq n} \delta({\bar \lambda}^{m}) \Omega^{n}_{m} \mathcal{F}^{m}}$}, with the $\mathcal{F}^{n}$ being arbitrary functions on the Berry curvature modified energy shells. If we now define, in all generality, the functions $\smash{{\bar f}^{n}(x, {\bf p})}$ through \mbox{$\smash{\mathcal{F}^{n} = 2 {\bf i} \pi \left(1 - \frac{\hbar}{2} \Omega^{n}_{n} \right) {\bar f}^{n}}$}, and ignoring some $O(\hbar^2)$ terms, we obtain: 
\begin{equation} \label{eq:lett_less_out_eq}
{\bar g}^{n}_{<} = 2 {\bf i} \pi \left[ \left( 1 - \frac{\hbar}{2} \Omega^{n}_{n} \right) {\bar f}^{n} \delta({\bar \lambda}^{n})  +  \frac{\hbar}{2} \sum_{m \neq n} \Omega^{n}_{m} {\bar f}^{m} \delta({\bar \lambda}^{m}) \right] , 
\end{equation}
which reduces to $\smash{{\bar g}^{(eq) n}_{<}}$, as given by Eq.(\ref{eq:lett_less_eq}), under the equilibrium boundary condition:
\begin{equation}\label{eq:lett_eq_BC}
{\bar f}^{(eq) n} = n_F({\bar \varepsilon}^{n}) . 
\end{equation}
This allows to identify $\smash{{\bar f}^{n}}$ as the out-of-equilibrium semiclassical occupation number functions. For a more straightforward comparison with Ref.\cite{wong:2011}, we will now specialize our discussion for the rest of this work to a general two-band Hamiltonian. We then have $\smash{\underline{\underline{\Omega}}^{\sigma} _{\sigma}{} = \underline{\underline{\Omega}}^{\sigma}_{-\sigma}{} = \sigma \underline{\underline{\Omega}}}$, $\smash{ \Delta^\sigma = \sigma \Delta }$ and \mbox{${\bf b}_{\sigma} = \sigma {\bf u}$,} with $\smash{ \sigma \in \lbrace -1 , 1 \rbrace \equiv \lbrace \downarrow , \uparrow \rbrace }$ and ${\bf u} \in {\mathbb R}^3$ with \mbox{${\bf u} \cdot {\bf u} = 1$.}  In order to get a governing equation for the ${\bar f}^{\sigma}$, we consider now the QKE, which yields:
\begin{eqnarray} \label{eq:lett_semi_diff}
& \left \lbrace {\bar \lambda}^{\sigma},  {\bar g}^{\sigma}_{<} \right \rbrace - \sigma \frac{\hbar}{2} {\bf u} \cdot \left( \left \lbrace {\bar \lambda}^{\sigma},  {\bf u} \right \rbrace \times \left \lbrace {\bar g}^{\sigma}_{<}, {\bf u}  \right \rbrace \right) & \nonumber \\& = - \hbar \left[ \left \lbrace \Delta ,  {\bar g}^{-\sigma}_{<} \right \rbrace \Omega + \frac{1}{2} \left \lbrace {\bar \lambda}^{\sigma} ,  \Omega \right \rbrace {\bar g}^{s}_{<} \right] + O(\hbar^2), &
\end{eqnarray} 
in which we have introduced $\smash{{\bar g}^{s}_{<} = {\bar g}^{\uparrow}_{<} - {\bar g}^{\downarrow}_{<}}$. This (collisionless) quantum Boltzmann equation, is the Green's function  counterpart of the density matrix evolution equation first derived by Wong and Tserkovnyak {\em i.e.}, Eqs.(19) and (21) in \cite{wong:2011}. In particular, we recover on the right-hand-side the same interband couplings seemingly at variance with the wave packet approach. However, substituting in Eq.(\ref{eq:lett_semi_diff}) our general solution to the CE {\em i.e.}, Eq.(\ref{eq:lett_less_out_eq}), the bands decouple and the QBE reduces into a remarkably compact and symmetric quantum Vlasov equation:
\begin{equation} \label{eq:lett_semi_diff2}
2 {\bf i} \pi \delta({\bar \lambda}^{\sigma}) \left[  \left\lbrace {\bar \lambda}^{\sigma},  {\bar f}^{\sigma} \right\rbrace - \sigma \frac{\hbar}{2} {\bf u} \cdot \left( \left \lbrace {\bar \lambda}^{\sigma},  {\bf u} \right \rbrace \times \left \lbrace {\bar f}^{\sigma}, {\bf u}  \right \rbrace \right) \right] = 0 .
\end{equation} 
The manifest elimination of all interband couplings resolves the previously found structural discrepancy between the wave packet and density matrix derived kinetic equations. This was not uncovered in \cite{wong:2011}, due to the absence of a constraint equation at the density matrix level. After expansion of the Poisson's brackets and energy integration, we obtain the Boltzmann-Vlasov equations in gauge invariant form for a general two-band system, with all leading order quantum geometric corrections:
\begin{subequations}
\begin{eqnarray}
&&\!\!\!\!\!\!\!\!\!\!\!\!\!\!\!\!\!\! 0  = \partial_t {\bar f}^{\sigma} + \left( \frac{d x^{i}}{d t} \right)^{\sigma}  \partial_{i}  {\bar f}^{\sigma} 
+ \left( \frac{d p_{i}}{d t} \right)^{\sigma} \partial_{p}^{i}  {\bar f}^{\sigma},  \label{eq:lett_QBE} \\
&&\!\!\!\!\!\!\!\!\!\!\!\!\!\!\!\!\!\!\!\!  (v^i)^\sigma  = \!\! \left( \frac{d x^i}{d t} \right)^{\sigma} \!\!\!\! =  -\partial_{p}^{i} \bar{\varepsilon}^{\sigma} + \sigma \hbar \left( \Omega^{j i} D_{j}  - \Omega_{\mu}{}^{i}{} \partial_{p}^{\mu} \right) {\lambda}^{\sigma}, \label{eq:lett_EOM_pos} \\
&&\!\!\!\!\!\!\!\!\!\!\!\!\!\!\!\!\!\!\!\! (F_i)^\sigma  = \!\! \left( \frac{d p_{i}}{d t} \right)^{\sigma} \!\!\!\! = \ \ \partial_{i} \bar{\varepsilon}^{\sigma} + e \left( c F_{i 0} + F_{j i} \partial_{p}^{j} \bar{\varepsilon}^{\sigma} \right) \nonumber\\
&& \ \ \ \ \ \ \ \ \ \ \ \ - \sigma \hbar \left[ \left( \Omega^{j}{}_{i}{} D_{j}  - \Omega_{\mu i} \partial_{p}^{\mu} \right)\right. \nonumber\\
&& \ \ \ \ \ \ \ \ \ \ \ \ + \left. (e/2) F_{j i} \left( \Omega^{k}{}^{j}{} D_{k}  - \Omega_{\mu}{}^{j}{} \partial_{p}^{\mu} \right) \right] {\lambda}^{\sigma}. \label{eq:lett_EOM_mom}
\end{eqnarray}
\end{subequations}
Our Eqs.(\ref{eq:lett_EOM_pos}-\ref{eq:lett_EOM_mom}) are found compatible with $O(\hbar)$ accuracy with the equations of motion derived with the wave packet approach \cite{sundaram:1999, xiao:2010}, up to quadratic order in an external constant and uniform field \cite{morimoto:2016}, hence providing to the later a confirmation of their validity within our framework. However, it has to be understood that in our theory these equations does not constitute single particle equations of motion independently derived, but are {\em definitions} of the kinetic velocity ${\bm v}$ and generalized force ${\bm F}$, as imposed by the derived kinetic equation {\em i.e.}, Eq.(\ref{eq:lett_QBE}).

{\em Expectation value of observables--} We shall now recall that a general expression for the diagonal part of the density matrix can be obtained within the wave packet approach, by combining Eq.(4) and Eq.(14) of Ref.\cite{xiao:2005}. Using our notations, it reads:
\begin{equation} \label{eq:lett_diag_density_wp}
{\bar \rho}_{W \! P}^{\sigma} =  \left( 1 - \sigma \hbar \Omega \right) {\bar f}^{\sigma} .
\end{equation}
The same quantity is obtained within our theory by energy integration of Eq.(\ref{eq:lett_less_out_eq}), which in the two-band case yields:
\begin{equation}
{\bar \rho}^{\sigma} = \bigg[ \left( 1 - \sigma \frac{\hbar}{2} \Omega\right) {\bar f}^{\sigma} + \sigma \frac{\hbar}{2} \Omega \ {\bar f}^{-\sigma} \bigg] .\label{eq:lett_diag_density_two_band}
\end{equation}
In addition, in the course of the derivation of Eq.(\ref{eq:lett_semi_diff}), the transverse part of the semiclassical density matrix, a quantity not provided by the standard wave packet approach, has also been obtained and it reads:
\begin{equation} \label{eq:lett_wig_trans}
\!\!\! {\bar {\bm \rho}}^{\bot} = - \frac{\hbar}{2} {\bf u} \times \!\! \left[ \lbrace {\bar \lambda}^0, {\bf u} \rbrace \frac{\left( {\bar f}^{\uparrow} - {\bar f}^{\downarrow} \right)}{\Delta} + \lbrace \left({\bar f}^{\uparrow} + {\bar f}^{\downarrow} \right), {\bf u} \rbrace \right] .
\end{equation}
Equations (\ref{eq:lett_diag_density_two_band}) and (\ref{eq:lett_wig_trans}) provide a complete representation of the semiclassical density matrix, according to ${\hat {\bar \rho}} = \sum_{\sigma} {\bar \rho}^{\sigma} {\hat P}_{\sigma} + \frac{1}{2} {\bar {\bm \rho}}^{\bot} \cdot \hat{\bm \sigma}$. This constitutes, together with Eqs.(\ref{eq:lett_QBE}-\ref{eq:lett_EOM_mom}), the central result of this work. The wave packet approach is not providing such a complete and $O(\hbar)$ accurate representation of the density matrix, notwithstanding our validation of its equations of motion. In order to further illustrate the physical consequences of this, we derive now the (dimensionless) charge  density $\langle \rho^c \rangle$ and (possibly pseudo) spin density $\langle {\bm \rho}^{s} \rangle$ in a general out-of-equilibrium situation.  First, following the standard prescriptions of the wave packet approach {\em i.e.}, using Eq.(\ref{eq:lett_diag_density_wp}) from this letter, and Eq.(6) of Ref.\cite{xiao:2005}, we obtain:
\begin{subequations}
\begin{eqnarray}
\langle \rho^c_{W\!P} \rangle = & & \sum_{\sigma} \int \frac{dp}{(2 \pi \hbar)^d} \left( 1 - \sigma \hbar \Omega\right) {\bar f}^{\sigma} , \\
\langle {\bm \rho}^{s}_{W\!P}  \rangle = & & \int  \frac{dp}{(2 \pi \hbar)^d} \left( {\bar f}^{\uparrow} -{\bar f}^{\downarrow} \right) {\bf u} \nonumber \\
& - & \hbar  \int  \frac{dp}{(2 \pi \hbar)^d} \Omega \left( {\bar f}^{\uparrow} + {\bar f}^{\downarrow} \right) {\bf u} . \label{eq:lett_spin_acc_wp}
\end{eqnarray}
\end{subequations}
Conversely, and in accordance with our Eqs.(\ref{eq:lett_diag_density_two_band}), (\ref{eq:lett_wig_trans}) and (\ref{eq:lett_wigner_avg}), we obtain:
\begin{subequations}
\begin{eqnarray}
\langle \rho^c \rangle & = & \sum_{\sigma} \int \frac{dp}{(2 \pi \hbar)^d} \left( 1 - \sigma \hbar \Omega\right) {\bar f}^{\sigma} , \label{eq:lett_char_dens}\\
\!\!\!\!\!\!\!\!\!\!\!\!\!\!\!\!\!\!\!\!\!\!\!\! \langle {\bm \rho}^{s} \rangle & = &  \!\!\!\!  \int  \frac{dp}{(2 \pi \hbar)^d} \left( {\bar f}^{\uparrow} - {\bar f}^{\downarrow} \right) {\bf u} + \int  \frac{dp}{(2 \pi \hbar)^d} {\bar {\bm \rho}}^{\bot} .\label{eq:lett_spin_acc}
\end{eqnarray}
\end{subequations}
The agreement for the charge density stems from the wave packet assumption \cite{xiao:2005} of a band projected density matrix satisfying the Liouville theorem in each band taken separately. This is not true in general, but it implies the weaker but generally true condition fulfilled by our theory, that the trace over the bands satisfies it with $O(\hbar)$ accuracy. However, the absence in our Eq.(\ref{eq:lett_spin_acc}) of the spurious second term appearing in Eq.(\ref{eq:lett_spin_acc_wp}) demonstrates the need to use our Eq.(\ref{eq:lett_diag_density_two_band}) when computing the expectation value of general observables. This is in addition to the needed consideration of a possible transverse contribution \cite{wickles:2013,sekine:2017}, as seen in our Eq.(\ref{eq:lett_spin_acc}) for the spin density, which is also absent from the standard wave packet approach.   

{\em Applications--} To concretely demonstrate the efficiency of our formalism in capturing the leading order quantum geometric corrections to electron transport properties, we focus now on the case of an Hamiltonian with no explicit spacetime dependence. We first express the kinetic velocity as resulting from our Eq.(\ref{eq:lett_EOM_pos}), and the effective force as resulting from our Eq.(\ref{eq:lett_EOM_mom}), in explicit vector form in this limit case. We have:
\begin{subequations}
\begin{eqnarray}
	{\bm v}^\sigma &=& \partial_{\bf p} \bar{\varepsilon}^{\sigma} \nonumber \\  
	&-& e \hbar \sigma {\bm E} \times {\bm \Omega} \nonumber \\
	&+& e \hbar \sigma {\bm \Omega} \times \left( \partial_{\bf p} {\varepsilon}^{\sigma} \times {\bm B} \right) , \label{eq:wp_vel2}\\
	{\bm F}^\sigma &=& e \left( {\bm E} + \partial_{\bf p} \bar{\varepsilon}^{\sigma} \times {\bm B} \right) \nonumber \\
	&+& e^2 \hbar \sigma \left( {\bm B} \cdot {\bm \Omega} \right) \ \partial_{\bf p} {\varepsilon}^{\sigma} \times {\bm B}   \nonumber \\
	&-& e^2 \hbar \sigma \left( {\bm E} \times {\bm \Omega} \right) \times {\bm B}  .	\label{eq:wp_for2}
\end{eqnarray}
\end{subequations}
The Boltzmann-Vlasov equation {\em i.e.}, Eq.(\ref{eq:lett_QBE}), can now be expressed more classically as:
\begin{equation}
\partial_t {\bar f}^{\sigma} + {\bm v}^\sigma \cdot \partial_{\bf x} {\bar f}^{\sigma} + {\bm F}^\sigma \cdot \partial_{\bf p}  {\bar f}^{\sigma} = 0.
\end{equation}
Performing a linear combination between the two bands in order to get to the charge density in accordance with Eq.(\ref{eq:lett_char_dens}), and after momentum integration, we obtain:
\begin{align} \label{eq:curr_cont1}
	& \ \ \ \ \partial_t \rho^{c} \nonumber \\
	&+ \partial_{\bf x} \cdot \sum_{\sigma} \int_{BZ} \frac{dp}{(2 \pi \hbar)^{d}} \left( e Z^{\sigma}_{c} \bar{f}^\sigma  \right) {\bm v}^\sigma \nonumber \\ 
	&+ \sum_{\sigma} \int_{BZ} \frac{dp}{(2 \pi \hbar)^{d}} Z^{\sigma}_{c}  {\bm F}^\sigma \cdot \partial_{\bf p} \left(e \bar{f}^\sigma  \right) = 0 ,
\end{align}
in which we recall that \smash{$Z^{\sigma}_{c} = 1 - e \hbar \sigma {\bm B} \cdot   {\bm \Omega}$}, and with all the integrals being over the first Brillouin zone (BZ). One can then easily verify that:
\begin{equation}
	\partial_{\bf p} \cdot \left( Z^{\sigma}_{c}  {\bm F}^\sigma \right) = - e^2 \hbar \sigma  \left( {\bm E} \cdot {\bm B} \right) \partial_{\bf p} \cdot  {\bm \Omega} ,
\end{equation}
which, by virtue of elementary vector calculus identities and uses of the divergence theorem, allows us to transform Eq.(\ref{eq:curr_cont1}) into the following equivalent form:
\begin{equation} \label{eq:curr_cont12}
	\partial_t \rho^c + \partial_{\bf x} \cdot {\bm j}^c =  - e^3 \hbar \sigma  {\bm E} \cdot {\bm B} \sum_{\sigma} \int_{FS}  \frac{dp}{(2 \pi \hbar)^{d-1}}  {\bm n} \cdot  {\bm \Omega} ,
\end{equation} 
which we recognize as the current continuity equation. The right hand side (RHS), which presents itself as proportional to the outgoing flux of the Berry curvature vector through the Fermi surface (FS), with ${\bf n}$ the unit normal vector at the FS, corresponds to the condensed matter counterpart \cite{sekine:2017} of the chiral (or triangle) anomaly of relativistic quantum field theory \cite{adler:1969}. The charge current ${\bm j}^c$ is found equal to:
\begin{align} \label{eq:char_curr}
{\bm j}^c = \ \ \ \ \ \ \  e &\sum_{\sigma} \int_{BZ} \frac{dp}{(2 \pi \hbar)^{d}} {f}^\sigma \partial_{\bf p} {\varepsilon}^{\sigma} \nonumber \\
			 - e^2 \hbar &\sum_{\sigma} \sigma {\bm E} \times \int_{BZ} \frac{dp}{(2 \pi \hbar)^{d}} {f}^\sigma  {\bm \Omega} \nonumber \\
			 - e^2 \hbar &\sum_{\sigma} \sigma \int_{BZ} \frac{dp}{(2 \pi \hbar)^{d}} {f}^\sigma \left( \partial_{\bf p} {\varepsilon}^{\sigma} \cdot {\bm \Omega} \right) {\bm B} .
\end{align} 
The first line term on the RHS corresponds to the ``normal'' current, which can have a finite value only for an anisotropic distribution function driven by an electric field. Conversely, we see that even with ${f}^\sigma$ made equal to the Fermi-Dirac distribution function \smash{$n_F({\varepsilon}^{\sigma})$}, both the second and third line terms on the RHS yield possibly nonzero currents at first order in the field. This corresponds to the intrinsic anomalous Hall effect (AHE) for the second line term, and to the chiral magnetic effect (CME) for the third line term. We then immediately obtain from Eq.(\ref{eq:char_curr}) the universal expressions for the intrinsic AHE conductivity:
\begin{equation}\label{eq:AHE_cond}
\sigma^{x y}_{\sigma} = -  \frac{\sigma e^2}{\hbar} \int \frac{d^2p}{(2 \pi)^2 \hbar^{(d-2)}} n_F\left[ \varepsilon^{\sigma}({\bf p}) \right] \Omega^{x y}({\bf p}) ,
\end{equation}
and CME coefficient:
\begin{equation} \label{eq:CME_coeff}
\alpha_{\sigma} = - \frac{\sigma e^2}{\hbar^2} \int \!\! \frac{d^3p}{(2\pi)^3 \hbar^{(d-3)}} n_F\left[ \varepsilon^{\sigma}({\bf p}) \right]\partial_{\bf p}\varepsilon^{\sigma}({\bf p}) \cdot {\boldsymbol \Omega}({\bf p}) ,
\end{equation}
for the band $\sigma$ of a completely general 2-band Hamiltonian. Hence, getting the actual value of these linear response coefficients for any particular model Hamiltonian reduces into computing the relevant Berry curvature tensor component in momentum space. For instance, in the case of a massive Dirac model with: 
\begin{equation}\label{eq:ham_DM}
\hat{h} = v_F {\bf p} \cdot {\bm \sigma} + \delta \hat{\sigma}_{z} ,
\end{equation} 
one can immediately obtain:
\begin{equation}\label{eq:2D_BC_mom}
\Omega^{x y}({\bf p}) = - \frac{v_F^2 \delta}{2 \left(v_F^2 p^2 + \delta^2 \right)^{3/2}} ,
\end{equation} 
by application of Eq.(\ref{eq:lett_mixed_gauge_berry}) to the momentum sector, matching previous calculations \cite{sinitsyn:2007}. Then the corresponding intrinsic AHE conductivity is obtained from Eq.(\ref{eq:AHE_cond}). Similarly, our general expression (\ref{eq:CME_coeff}) allows us to recover effortlessly the CME found in a two-band, 3D lattice model of Weyl semimetal {\em i.e.}, Eq.(20) in \cite{chang:2015}.

{\em Conclusion--} We have derived, for the first time as a controlled approximation of the gauge invariant Keldysh formalism, a complete semiclassical kinetic theory for multiband systems with non-trivial quantum geometry in the clean limit. We validate the previously proposed Berry curvature corrected semiclassical equations of motion. However, interband effects limit the relevance of a previously introduced correction to the density of states and modify the expectation value of observables. As an application, we have demonstrated how the current continuity equation with chiral anomaly can be straightforwardly recovered, readily providing general expressions for the AHE conductivity and the CME coefficient. The built-in parameterization of the quantum geometry in terms of Bloch vectors facilitates the evaluation of such transport coefficients for any specific model, through the computation of the relevant Berry curvatures. This illustrates how our theory delivers exact leading quantum corrections to transport properties while providing a valuable insulation from the complexity of the underlying quantum kinetic level. Furthermore, its rigorous foundation pave the way for future studies of the interplay between quantum geometry, field gradients, disorder, and interaction effects ; all important issues, some of which having been already investigated with the density matrix approach \cite{culcer:2017,sekine:2017,atencia:2022}. Finally and circling back to high energy physics which provided some of the technical tools used in this work, we would like to raise the interesting question of what are, if any, the relativistic quantum kinetic \cite{huang:2018,gao:2019,weickgenannt:2019,hidaka:2022} counterparts and physical consequences of the $O(\hbar)$ interband couplings found here at the non-relativistic level. 
\acknowledgements
The authors would like to acknowledge fruitful discussions with F. Pi\'echon,  Y. Tserkovnyak and X. Waintal as well as the financial support of the Dipartimento di Matematica e Fisica  Universit\`a  Roma Tre under the Visiting Eccellenza program.

\bibliography{berry}
\bibliographystyle{eplbib.bst}
\end{document}